# The Periodic Table as a Part of the Periodic Table of Chemical Compounds

Mikhail M. Labushev

*Department of Ore Deposits Geology and Prospecting, Institute of Mining, Geology and Geotechnology, Siberian Federal University, Russian Federation, mlabushev@yandex.ru*

The numbers of natural chemical elements, minerals, inorganic and organic chemical compounds are determined by 1, 2, 3 and 4-combinations of a set 95 and the numbers of combinations are respectively equal to 95, 4,465, 138,415 and 3,183,545. To explain these combinatorial relations it is suggested the concept of information coefficient of proportionality as mathematical generalization of the proportionality coefficient for any set of positive numbers. It is suggested a hypothesis that the unimodal distributions of the sets of information coefficients of proportionality for atomic weights of chemical elements of minerals and chemical compounds correspond to unimodal distributions of the above sets of combination of 2, 3 and 4 atomic weights of a set 95 natural chemical elements. The expected values of symmetrized distributions of information coefficients of proportionality sets for atomic weights of minerals and chemical compounds are proposed to be used to define chemical compounds, like atomic weights define chemical elements. Variational series of the expected values can be represented as a sequence of packets of 95 values. The Periodic Table should be extended to the Periodic Table of Chemical Elements and Chemical Compounds consisting of a short period with 95 chemical elements and 4,465 minerals, and 24 periods with 138,415 compounds each. All naturally occurred chemical elements, minerals and chemical compounds should be packed in 95 to compose the suggested Table. The first long period comprises the packets of inorganic compounds, the rest of the periods include the packets of organic compounds.

**INTRODUCTION**

After the discovery of the Periodic Law many attempts were made in search of the natural laws of molecular periodicity, the review of the researches is given in ref. [1, 2, 3]. In the 19th century chemists discussed the similarity of chemical elements and saturated hydrocarbons. N.A. Morozov tried to express this analogy in a form of a table. In his table hydrocarbon radicals of the homological hydrocarbons are arranged in columns the way similar elements are arranged in the Mendeleev Periodic Table [4]. Later the two-dimensional model with the numbers of carbon and hydrogen atoms on the axes was proposed by H. Decker [5].

H. G. Grimm formulated the Hydride Displacement Law (an early hypothesis to describe bioisosterism) as an ability of certain chemical groups to function as or similar to other chemical groups. The set of Hydride Displacement Law objects was presented in a two-axis coordinate system characterized by valence and the number of hydrogen atoms [6].

Important facet of molecular periodicity is connected with periodicity in sets of diatomic molecules. These molecules are the simplest case of polyatomic molecules but there is no simplicity in their periodic classifications and many scientists who studied them never looked at their results with respect to molecular periodicity [2]. Periodic tables of diatomic molecules were created on the base of the periodic change of spectroscopic constants[7]. Periodicity of properties of organic molecules is described by their systematics of the spectral-luminescent properties [8].

To characterize the diatomic molecules there were used the period numbers of two atoms in the Mendeleev's Periodic Table [9, 10], the row number as the sum of the period numbers of the atoms [9,10] and the same descriptions for triatomic molecules[11], the number of electrons for each atom[9,10], molecular orbital configurations [10, 12], data for dissociation potentials [10]. V.M. Kaslin proposed 44 periodic tables of diatomic molecules (the atoms of the chemical groups of lithium, beryllium, boron and carbon) on the basis of charges of the nuclei of atoms, atomic weight and atomic parameter called molecular affinity [13, 14].

New approaches to molecular periodicity in the 1980s were considered by A. Haas [15, 16, 17], J. Dias [18, 19] and A.P. Monyakin [20]. A. Haas proposed the element displacement principle to explain the halogen-like properties of the pseudohalogen groups. It also allows the correlation of classical pseudohalogens with perfluoroorgano-element groups. Radicals derived in this way have the same valence-electron structure as the reference element and are termed paraelements. The paraelements can be arranged into a set of periodic tables. Haas proved analogies between the elements and paraelements by means of physical data, structural features and examples of their chemical similarity.

Dias described regularities for benzenoid polycyclic aromatic hydrocarbons using their periodic table. To construct the table the numbers of carbon and hydrogen atoms and "net number of disconnections among the internal edges" were used. The table reflects similarity relations among polyaromatic hydrocarbons. The ground of the periodic table is group theory.

Monyakin's periodic system has three axes that characterize separate atoms period numbers and combined group numbers [3].

The ideas of extended molecular classifications are connected with researches of P. Gorsky [21,22], R.A. Hefferlin, G.V. Zhuvikin, K.E. Caviness, P.J. Duerksen [23], E.V. Babaev [2] and R.A. Hefferlin [23, 24] . Since 1970s A. Gorsky has been developing the "morphological" classification of chemical compounds which is based on the atomic nucleus concept taking into account nucleus number, nuclear charge, electron shells and the number of valence sub-shells.

R.A. Hefferlin and others suggested periodic systems of N-atom molecules. By means of orthonormal transformations molecules can be arranged according to the numbers of electrons being different in atomic numbers of their constituent atoms.

E.V. Babaev suggested using the concept of hyperperiodicity to solve the problem of classification of isovalent ensembles of molecules with their arrangement on the plane "the number of valent electrons- the number of atoms" by adding three axes to the plane, i.e. some sort of space with five dimensions. The third axis is defined by parameter which can be sensitive to the number of inner shells (or inner electrons) of atoms and two additional axes haven't been defined yet.

It must be separately mentioned the search for the molecular periodicity having been held by Southern Adventist University (SAU) group. These researches were organized by R. Hefferlin, who associates the results of his researches of molecular periodicity with the work of the group as well. The SAU group has conceived as a paradigm the construction of periodic systems for molecules with any number of atoms. Theoretical support is provided by reducible representations in the Hilbert spaces of molecules. Included in the paradigm is the concept that atoms maintain enough of their individualities to their "own" valence-shell electrons [24].

None of the proposed models of molecular periodicity is based on the atomic weights (relative atomic masses) of chemical elements. It should be noted, that it is the very factor that has made it possible to develop the Periodic Table of Chemical Elements. The longtime research of molecular periodicity has failed to show the place of all or part of the known chemical compounds in the general periodic system.

Dmitri Mendeleev's main goal was to find factors determining properties not only of simple bodies but also of chemical compounds. As he formulated it in 1871 and 1879: «The properties of simple bodies, the constitutions of their compounds, as well as the properties of these last, are periodic functions of the atomic weights of the elements.»

**A NEW FACTOR FOR CHEMICAL COMPOUNDS CHARACTERIZING**

Can we receive any universal factor for chemical compounds similar to atomic weight for chemical elements? If it is possible, it should be evaluated on the basis of atomic weights of chemical elements and shouldn't vary at their proportional reduction or increase. Proportionality coefficient satisfies these requirements, but for the calculations it's necessary to get its mathematical generalization.

It is accepted to consider proportionality to be elementary relation and proportionality coefficient is no more than quotient of two numbers, therefore, it can be computed only for two atomic weights. In this case, there is an uncertainty associated with the choice of dividend and divisor for calculating proportionality coefficient. As a result of the numbers 1 and 2 proportionality coefficient calculations we obtain 0.5 and 2. Which of the two coefficients should characterize the proportionality of the numbers 1 and 2?

We hardly know how to determine the proportionality of two numbers while in nature the proportionality of three numbers is more important possibly due to three-dimensionality of space. For example, a human eye has three types of color sensors that respond to different ranges of wavelengths, a full plot of all visible colors is a three-dimensional figure. No wonder they say that God loves a trinity!

The number of natural chemical elements, minerals, inorganic and organic compounds is approximated by the number of combinations of 1, 2, 3 and 4 in a set of 95, being respectively equal to 95, 4465, 138,415, 3,183,545 as binomial coefficients and all of them being multiples of 95.

These relationships require explanation, as, for example, the presence of two atoms of two chemical elements in a chemical formula of a mineral (such as NaCl) is neither necessary nor sufficient condition. Formula of mineral can contain up to 12 chemical elements, like for mineral hiortdahlite $Na_4Ca_8Zr_2(Nb,Mn,Ti,Fe,Mg,Al)_2(Si_2O_7)_4O_3F_5$.

To explain these relations it is suggested the concept of information coefficient of proportionality $I_p$ as mathematical generalization of the proportionality coefficient for any set of positive numbers [25, 26]. It is suggested a hypothesis that the unimodal distributions of $I_p$ sets for atomic weights of chemical elements of minerals and chemical compounds correspond to unimodal distributions of $I_p$ sets for combination of 2; 3 and 4 atomic weights of 95 natural chemical elements.

The expected values $I_{av}$ of symmetrized distributions of $I_p$ sets for atomic weights of minerals and chemical compounds are proposed to be used to define chemical compounds, like atomic weights define chemical elements. Variational series of $I_{av}$ can be represented as a sequence of packets of 95 values.

The structure of each packet can be characterized by seven periods like seven periods characterize D. Mendeleev's Periodic Table of chemical elements. It is assumed that the properties of chemical compounds in each packet vary similar to the properties of chemical elements in the Periodic Table. In any two adjacent packets relations $I_{av}$ of the same rank differ on value of a constant being estimated as 1.000013.

The number of naturally occurring chemical elements and minerals is known to be the most reliable data. The information on minerals can be obtained from the International Mineralogical Association (IMA) Database of Mineral Properties which is available on the rruff website http://rruff.info/ima/#. The number of known organic and inorganic chemical compounds can be estimated by the collection of chemical compounds of the National Institutes of Health, which is expected to contain 3,000,000 compounds.

It should be taken into account that the IMA list includes some native chemical elements as well. If we assume that all first 95 chemical elements may occur in nature in crystalline state, the total number of minerals should be taken as 4,560. On March 15, 2011 the list includes 4,503 known minerals, and 4,560 ones happen to approximate above this number.

**METHODS**

In information theory, entropy is a measure of uncertainty associated with a random variable. The entropies H(X) and H(Y) of discrete random variables X and Y with possible values $\{x_1, ..., x_n\}$ and $\{y_1, ..., y_m\}$ are:

$$H(X) = -\sum_{i=1}^{n} p(x_i) \log_b p(x_i),$$

$$H(Y) = -\sum_{j=1}^{m} p(y_j) \log_b p(y_j),$$

where b is the base of the logarithm used. Common value of b is 2, Euler's number e, and 10, let b equal 10 in our case.

The mutual information of two random variables is a quantity that measures the mutual dependence of the two variables. Mutual information can be expressed as

$$I(X,Y) = H(X) + H(Y) - H(X,Y).$$

where H(X,Y) is the joint entropy of X and Y:

$$H(X,Y) = -\sum_{i=1}^{n}\sum_{j=1}^{m} p(x_i, y_j) \lg p(x_i, y_j)$$

Information coefficient of proportionality $I_p$ is suggested being calculated with similar formulas. Let X and Y be dependent random variables with values $\{x_1, ..., x_n\}$ and $\{y_1, ..., y_m\}$ being presented as a matrix consisting of m rows and n columns in which the elements of every row and every column are the same, n=m, $x_{ij}=y_{ij}$ and number of different $x_{ij}$ is equal to n. Ecuation $x_{ij}=y_{ij}$ is satisfied for all values of the involved variables, thus it is an identity relation. This case is not interesting for the calculation of mutual information between X and Y, but makes it possible to define the proportionality coefficient for all $x_{ij}$.

In $I_p$ calculation probabilities are replaced by information coefficients of string, column and matrix proportionality K (X), K (Y) and K (X, Y).

$$K(X) = -\sum_{i=1}^{n} k(x_i) \lg k(x_i),$$

$$K(Y) = -\sum_{j=1}^{m} k(y_j) \lg k(y_j),$$

$$K(X,Y) = -\sum_{i=1}^{n}\sum_{j=1}^{m} k(x_i, y_j) \lg k(x_i, y_j),$$

$I_p = K(X) + K(Y) - K(X,Y)$.

K(X), K(Y) и K(X,Y) are calculated using the coefficients of row, column and matrix proportionality k:

$$k(x_i) = \frac{\sum_{i=1}^{n} x_i}{\sum_{i=1}^{n}\sum_{j=1}^{m} x_{ij}},$$

$$k(y_j) = \frac{\sum_{j=1}^{m} y_j}{\sum_{i=1}^{n}\sum_{j=1}^{m} y_{ij}},$$

$$k(x_i, y_j) = \frac{x_{ij}}{\sum_{i=1}^{n}\sum_{j=1}^{m} x_{ij}}.$$

The computer simulated results have been obtained with a special case when the number of different values of X and Y is equal to three, n=m=3, $x_1=y_1$, $x_2=y_2$, $x_3=y_3$. Let these numbers compose a matrix of order 3*3 in which the elements of every row and every column are the same. For numbers 2, 3 and 4, one of the matrices can be summarized as follows: first, second and third rows of the matrix contain numbers in the following sequences: (2, 3, 4), (3, 4, 2) and (4, 2, 3). It allows calculating $I_p$ for three of these numbers uniquely. Similar calculations with n = m and n> 3 make it possible to calculate the values of $I_p$ as well, but further mathematical operations on them are admissible only under the condition n = const for all calculated $I_p$.

Proportionality of any quantity of numbers under the condition n = m = 3 can be characterized by calculation of a set of $I_p$ values for random subsets of initial set of numbers, the subsets containing 8 input numbers each, the ninth number in each subset is calculated as the sum of the remaining eight numbers and is called the total element. This scheme eliminates the problem of the homogeneity of $I_p$ sets and makes it possible to determine $I_p$ for one number as a constant equal to 0.04782, which is suggested being called information constant of monoelement proportionality. Thus can be calculated $I_p$ for one and two numbers (Fig.1).

| 1,008 | 1,008 | 1,008 |
|---|---|---|
| 1,008 | 8,064 | 1,008 |
| 1,008 | 1,008 | 1,008 |

a

| 1,008 | 4,003 | 1,008 |
|---|---|---|
| 4,003 | 1,008 | 1,008 |
| 4,003 | 4,003 | 20,042 |

b

| 4,003 | 1,008 | 1,008 |
|---|---|---|
| 23,037 | 4,003 | 1,008 |
| 4,003 | 4,003 | 4,003 |

c

| 1,008 | 1,008 | 1,008 |
|---|---|---|
| 4,003 | 1,008 | 17,048 |
| 1,008 | 4,003 | 4,003 |

d

**Figure 1**: Shows the cases of $I_p$ calculation for atomic weight of hydrogen (a) and atomic weights of hydrogen and helium (b, c, d). In the first case $I_p$ is the information constant of monoelement proportionality. In the second case, the atomic weights of hydrogen and helium appear 12 times in the calculation of a set of three $I_p$. Probabilities of their selection for calculations correspond to probabilities of their random selection among the initial data. These calculations are insufficient for the description of the two indicated atomic weights proportionality, because many variants are not presented, for example, those including 2 and 6, 7 and 1 atomic weights of each of the two chemical elements with different distribution in the matrixes and the probability of all variants occurrence is not taken into account.

Proportional change of all numbers does not influence the quantity of calculated $I_p$, the property being possessed by common coefficient of proportionality as well. For calculating the parameters of the received distribution of a set of $I_p$ values with given accuracy, Ip set should be sufficiently large and calculated in compliance with the following conditions:
- relative frequencies of using numbers to calculate coefficients of row, column and matrix proportionality k can vary from calculation to calculation but the probabilities must be equal to the probabilities of their random selection from the initial numbers without taking into account the total elements;
- any $x_{ij}$ must be bigger than 0;
- the distribution of initial numbers in matrix is random.

**CHARACTERISTICS OF DISTRIBUTIONS OF $I_P$ SETS FOR ATOMIC WEIGHTS**

About 300 distributions of $I_p$ sets have been evaluated. These are the distributions for atomic weights of chemical elements in minerals, rocks, periods and groups of Mendeleev's Periodic Table of Chemical Elements and for the systems of atomic weights: atomic weight of hydrogen, atomic weights of hydrogen and helium, atomic weights of hydrogen, helium, ..., has-sium with the successive increase in their composition from one to 108 atomic weights in ascending order of atomic number of a chemical element.

In all cases these distributions are unimodal right-asymmetric being symmetrical more rarely, they can be symmetrized satisfactorily by square rooting from $I_p$ values or by taking the logarithms. In all cases there has been determined with Pearson criterion the significant difference of $I_p$ sets distributions from the normal and lognormal laws at confidence level 0.05 and number of degrees of freedom 4. There haven't been received any pair of equal values of $I_{av}$ for various systems of atomic weights.

Distributions of $I_p$ values in the systems of atomic weights: atomic weight of hydrogen, atomic weights of hydrogen and helium, atomic weights of hydrogen, helium, ..., hassium with the successive increase in their composition from one to 108 atomic weights in ascending order of atomic number of a chemical element, are unimodal symmetrical or right-asymmetric views being symmetrized satisfactorily by square rooting from $I_p$ values (Fig.2). These systems are suggested being called the additive systems of atomic weights of order n, where n equals the number of atomic weights in the system.

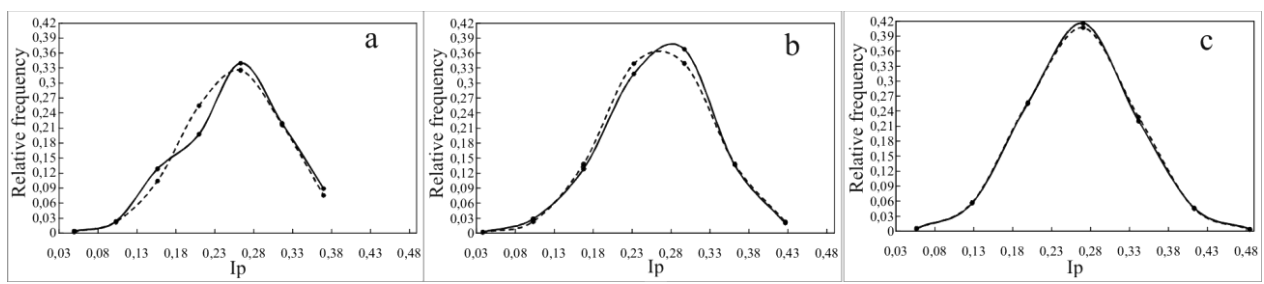

**Figure 2:** Symmetrized distributions of $I_p$ values in the additive systems of the atomic weights of the 2nd (a), 3rd (b) and 108th (c) orders. The dotted lines point out theoretical normal distributions for these sets, with the number of information coefficients of proportionality in each set being equal to 43,740. Mean values, standard deviations and empirical values of chi-square with four degrees of freedom are respectively equal to: a) 0.2565, 0.0642 and 1008.5720, b) 0.2650, 0.0657 and 279.3654, c) 0.2665, 0.0667 and 40.6921

The importance of proportionality of atomic masses of elements to describe systems of chemical elements was discovered while testing Gold Alloy State Standard Materials GSO 8754-2006 - GSO 8763-2006 produced by JSC «Krastsvetmet». They contain gold, silver and copper. Krastsvetmet Analytical Laboratory carried out homogeneity testing of the reference materials under the method specified by GOST 8.315-97.

For this purpose 25-50 articles of each reference materials were selected with their analytical surfaces being prepared. On each surface they were carried out two measurements of the chemical elements contents with X-ray fluorescence analysis without changing the position of the sample. After the measuring each article analyzed was cut along a plane parallel to the analytic surface. The position of the cut plane was chosen randomly throughout its length (height). As a result they were received from 100 to 200 analyzes of each reference material.

Variational series of chemical elements contents normalized to the median value were compared with normalized to the median value $I_{av}$ values for $I_p$ sets of additive systems of atomic weights of various orders. For these systems $I_p$ sets of more than 40,000 values were calculated with computer simulation being further symmetrized by square rooting. After that arithmetic means $I_{av}$ for symmetrized samples were calculated. It was ascertained the similarity of variational series of $I_{av}$ additive systems of atomic weights of 20th, 30th and 108th orders to normalized contents of various chemical elements (Fig.3) in the certified reference materials [27].

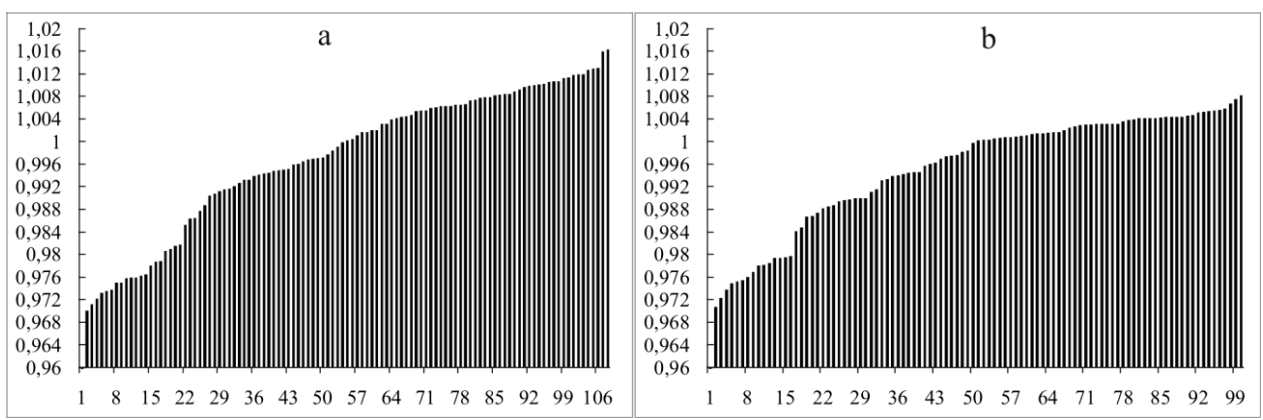

**Figure 3**: Variational series of normalized to the median value quantities: a) $I_{av}$ of additive systems of atomic weights of chemical elements from 1st to 108th orders, (b) copper contents in GSO 8758-2006. The first quantities are not shown in the histograms because of their small values equal to 0.8269 and 0.9589 respectively. This similarity is emphasized by the presence of typical minimum values of the studied quantities and the absence of values around 0.984 of both medians. In theoretical model minimum value is equal to normalized square root of information constant of monoelement proportionality.

To determine the best similarity with the variational series of normalized contents of copper of the fifth standard sample the variational series of $I_{av}$ of additive systems of atomic weights of 71st, 95th and 108th orders were considered. All these series are characterized by marks - $I_{av}$ values significantly higher than the previous one and approximately equal to the next $I_{av}$ in ascending order.

These marks are the most clearly visible at the diagrams in the fields of normalized quantities with values from 0.970 to 0.995 of the medians, and their equal amount has been registered in all variational series. The $I_{av}$ marks have been compared with the similar marks of normalized copper contents of GSO 8758-2006 (Tab.1).

**Table 1:** Marks of variational series of normalized values

| Marks of normalized copper contents | Marks of $I_{av}$ of additive systems of atomic weights | | | |
|---|---|---|---|---|
| | Order | | | The average for the 108th and 95th orders |
| | 108th | 95th | 71st | |
| 0.971 | **0.970** | 0.973 | 0.976 | 0.971 |
| 0.975 | **0.975** | 0.978 | 0.981 | 0.976 |
| 0.978 | 0.976 | **0.979** | 0.982 | 0.977 |
| 0.979 | **0.978** | 0.981 | 0.984 | 0.980 |
| 0.984 | 0.981 | **0.984** | 0.985 | 0.982 |
| 0.987 | 0.985 | **0.988** | 0.986 | 0.987 |
| 0.988 | 0.986 | **0.989** | 0.991 | 0.988 |
| 0.989 | **0.988** | 0.992 | 0.992 | 0.990 |
| 0.991 | **0.990** | 0.993 | 0.995 | 0.992 |

Note. The boldface font marks out theoretical values of $I_{av}$ for additive systems of 108th and 95th orders atomic weights closest to the respective normalized contents of copper.

The marks of variational series of $I_{av}$ of additive systems of atomic weights of the 71st order are the most different from the marks of the variational series of normalized copper contents. The marks of the remaining two rows of the variational series of $I_{av}$ complement each other; the average value of their $I_{av}$ gives the best approximation to the marks of normalized copper contents.

The analysis of variational series of the normalized contents of chemical elements of other reference materials has showed their similarity to variational series of $I_{av}$ of additive systems of atomic weights of chemical elements of the 20th and 30th orders.

**THE UNITED PERIODIC TABLE OF CHEMICAL ELEMENTS AND COMPOUNDS**

The creation of Periodic Table by D. Mendeleev showed the importance of studying the periodic changes of properties of chemical compounds in connection with atomic weights of chemical elements generating them. Later it was concluded that the properties of chemical ele-ments are in periodic dependence on the charge of atomic nuclei of chemical elements. But this does not explain any regularities related to isotopes distributions.

Some important natural processes cause changes in the isotopic composition of elements: e.g. the changes in groundwater before and during earthquakes. It is described in the discovery number 129 «The phenomenon of changing the chemical composition of groundwater by earthquake» made by G. Mavlyanov, V. Ulomov and others with priority from 21 February 1966 in the State Register of Discoveries of the USSR.

The formula of the discovery is the following one: «It was established a previously unknown phenomenon of changing the chemical composition of groundwater caused by earthquake, which lies in the fact that in the periods preceding earthquake, as well as during its process, the concentration of microelements, i.e. noble gases (radon, helium, argon), fluorine, uranium, increases and their isotopic composition changes in the groundwater, territorially connected with epicentral zone. »

Isotopic ratios are characterized by atomic weights and are not accompanied by the changes of charges of atomic nuclei of chemical elements, therefore, periodic dependence of the properties of chemical elements on the nuclear charge of atoms of chemical elements can't be used for their research.

Atomic nucleus charge does not determine the development of the periodic law, as there aren't any important theoretical forecasts on this basis. For example, it is impossible to determine the upper limit of the Periodic Table and forecast the number of possible chemical compounds. Some researchers already suggest the need of abandoning the periodic law for the further development of pure chemistry. To resolve this problem, it is suggested that the Periodic Table should be extended to the Periodic Table of Chemical Elements and Chemical Compounds with $I_{av}$ values for the description of chemical compounds.

It is suggested that $I_{av}$ for atomic weights of all the minerals and chemical compounds ranked in ascending order should be packed in 95 $I_{av}$. Minerals characterized by one packet indices should have similar and naturally changing physical and chemical properties. The same regularity is expected for inorganic and organic chemical compounds.

Packets compose the Periodic Table which consists of a short period with 95 chemical elements and 4,465 minerals, and 24 periods, each of them consists of 138,415 compounds. In the short period the first packet contains 95 chemical elements from hydrogen to americium, which show generality with minerals and under certain conditions must exist in crystalline state.

The first long period comprises packets of $I_{av}$ of inorganic compounds, the rest of the periods characterize organic compounds. The packets which are vertically adjacent in the table also seem to characterize chemical compounds which are similar in chemical and physical properties.

Individual packets can be presented in the form in which we are used to seeing the D. Mendeleev's Periodic Table.

The united Periodic Table of Chemical Elements and Chemical Compounds is based on the equality of distributions of $I_p$ sets calculated for atomic weights of 2 to 12 chemical elements that make up minerals and the distributions of $I_p$ sets for two, three or four atomic weights of the first 95 chemical elements. Computer simulation has confirmed the possibility of $I_{av}$ equality of twelve and two numbers. The approximate equality is established, for example, for $I_{av}$ of 46,500, 59,300, 45,650, 74,600, 54,400, 57,400, 40,500, 32,200, 35,400, 62,700, 41,500, 53,400 and atomic weights of aluminum 26.98154 and scandium 44.9559.

It is suggested a hypothesis that mathematical expectations of $I_{av}$ of symmetrized universes of $I_p$ of 95 minerals of the first packet, periodically increasing by proportionality constant, repeat 46 times. The same is expected for the 24 long periods of chemical compounds, each of them contains 1,457 packets. The basis for this hypothesis is the similarity of the normalized variational series of copper, silver and gold contents and normalized $I_{av}$ of additive systems of atomic weights of different orders. These systems are different in nature while displaying the amazing similarity of values variability.

The value of proportionality coefficient can be estimated. Minimum $I_{av}$ of symmetrized sets of $I_p$ is suggested that possess a value approximately equal to the square root of information constant of monoelement proportionality, and the maximum $I_{av}$ equal to 0.3447 has been computer simulated for the atomic weights of hydrogen, helium, lithium and plutonium. The distribution of $I_p$ for these atomic weights is expected to be reduced to a similar distribution of organic chemical compound which completes the periodic system. Based on this data the value of proportionality coefficient is estimated as 1.000013 for packets with 95 $I_{av}$ values.

If the proposed structure of the Periodic Table is accepted, the total number of organic compounds is multiple of the number of non-organic ones. These multiple ratios of number of combinations of 2, 3 and 4 for atomic weights from 89 to 118 are typical only for 107 atomic weights, and in this case the corresponding number of combinations is 5,671, 198,485 and 5,160,610. It is suggested for consideration the combination of two periodic systems based on 95 and 107 atomic weights with their certain resonance for the characterization of chemical compounds and probably mixtures as well.

There are 87 generally known minerals of mercury nowadays. They should probably be attributed to one packet. It is numbered to 81 minerals of ruthenium, rhodium and palladium. It is assumed that these minerals, as well as the minerals of mercury, form the basis for another packet. The total number of sulfur-containing minerals of bismuth is100, being probably almost all of them attributed to one packet. Minerals of light metals can also be the basis of the packets. For example 101 minerals of lithium without beryllium and 102 minerals of beryllium without lithium are generally known. It should be mentioned that there are about ninety generally known micas and micaceous minerals. The number of minerals of other light chemical elements is very large so it seems to be difficult to find their distribution in packets.

**DISCUSSION**

What is the importance of the mathematical generalization of the concept of proportionality coefficient? This can be assessed indirectly, by analogy with the concept of factorial. Gamma function is extends the concept of factorial function in the field of complex numbers. Chi-square distribution and Student's distribution being widely used probability distributions in mathematical statistics are related to Gamma function. Proportionality is not less important in the world than factorial ratios, but the proportionality coefficient is not extended to the case of three or more numbers and one number.

What are the preferences of $I_p$ calculation in comparison with H (X) or H (Y) entropies? Suggest initial numbers in the matrix and consider a case of an element commutation in the matrix. When the element is replaced in the row it's K (Y) that varies and the element replacement in the column changes K (X). While calculating the "mutual information" all cases of elements commutation in the matrix can be identified.

The similarity of variational series of $I_{av}$ of additive systems of atomic weights of chemical elements and contents of chemical elements in the alloys indicates the common basis of proportionality of atomic weights and clusters in the certified reference materials. This foundation can be qualitatively explained as follows.

Let us turn to the proportionality coefficients of atomic weights of hydrogen, helium and lithium. They are close to the proportionality coefficients of atomic weights of copper, silver and gold while being in the consideration together with the individual atoms, diatomic and tetra atomic clusters. This similarity is displayed through the comparison of normalized atomic weights (Tab.2).

**Table 2:** Initial and normalized atomic weights of chemical elements and clusters

| Chemical elements | Atomic weights | Normalized atomic weights* | Chemical elements and clusters | Atomic weights | Normalized atomic weights ** |
|---|---|---|---|---|---|
| H | 1.00794 | 1 | Cu | 63.546 | 1 |
| He | 4.002602 | 3.97 | CuAu | 260.51255 | 4.10 |
| Li | 6.941 | 6.89 | $Cu_2AgAu$ | 431.92675 | 6.80 |

Note. Atomic weights are normalized: * to atomic weight of hydrogen, ** to atomic weight of copper.

The number of listed platinoids is less than 95, and sulfur containing bismuth minerals can be attributed to one packet. This does not mean that there remain undiscovered three more minerals of these platinum group and three minerals of bismuth should be removed from the list of minerals. The result of studying the proportionality of atomic weights of rocks indicates the presence of at least two or three mineral-marks in each packet.

These minerals are interlinks between adjacent packets. In the Periodic Table of Chemical Elements there exist close analogues of such marks in a form of radioactive elements - technetium, promethium and polonium in the sequence of stable elements. The last of these elements unites (separates) radioactive and stable chemical elements.

Are there any analogues of the assumed periodicity properties of chemical elements and compounds in the nature? The short period is present both in the D. Mendeleev's Periodic Table and in the proposed system of chemical elements and compounds. It seems to be interesting the coincidence of the number of periods of the proposed periodic system with the number of hours in a day. Mean solar day is equal to 24 hours 3 minutes 56.554 seconds and can be represented as 24 long and one short period of time.

**CONCLUSION**

On the basis of the conducted research it is possible to forecast the number of undiscovered minerals. Since among 4,503 minerals certified by IMA by March 15, 2011 only 27 chemical elements are native crystal forms without taking into account 68 chemical elements of the first 95, which may exist in native crystalline form, but have not been found in nature yet.

The count of the number of minerals is complicated by the phenomena of polymorphism as the ability of certain chemical entities to crystallize in more than one form. We should also distinguish the polymorphism of chemical elements on the one hand from minerals as chemical compounds on the other one. For example, diamond, graphite, lonsdaleite and chaoite should be attributed to the chemical element carbon in crystalline form, i.e. they represent a chemical element and should be considered in the form of one chemical element, but not among 4,465 minerals. On the other hand, aragonite, calcite and vaterite with a single formula $CaCO_3$ in a set of 4,465 minerals correspond to only one mineral.

Computer simulation with atomic weights of chemical elements shows the new prospects of transition to a numerical characterization and modeling of physical and chemical properties of minerals, rocks and chemical compounds. There is every reason to launch international project «The Periodic Table of Chemical Elements and Compounds».

**ACKNOWLEDGEMENTS**

I appreciate sincerely Prof. R.A. Tsykin for comments and valuable advice. I also express gratitude to Prof. A.N. Gorban for the comments and recommendations. I gratefully acknowledge my wife E.I. Fomina for inspiration, editing and translating this article into English, my parents M.I. Labushev and T.V. Labusheva, and my sister T.M. Labusheva for funding my research.

materials for composition of base gold, OJSC "Krastsvetmet". *Certified Reference Materials* 4, pp. 52–64.